\documentstyle[psfig]{mn}

\begin{document} 
\def\lsim{\mathrel{\hbox{\rlap{\hbox{\lower4pt\hbox{$\sim$}}}\hbox{$<$}}}}
\def\gsim{\mathrel{\hbox{\rlap{\hbox{\lower4pt\hbox{$\sim$}}}\hbox{$>$}}}}
\def\simlt{\mathrel{\rlap{\lower 3pt\hbox{$\sim$}}
        \raise 2.0pt\hbox{$<$}}}
\def\simgt{\mathrel{\rlap{\lower 3pt\hbox{$\sim$}}
        \raise 2.0pt\hbox{$>$}}}

\title[The quasar epoch and the stellar ages of early-type galaxies]
{The quasar epoch and the stellar ages of early-type galaxies}

\author[A. Cattaneo \& M. Bernardi]
{A.~Cattaneo $^1$ \& M.~Bernardi $^2$\\
$^1$Institut d'Astrophysique de Paris, 98bis Boulevard Arago, 75014 Paris, France \\
$^2$Department of Physics, Carnegie Mellon University, Pittsburgh, PA 15213, USA}

\maketitle

\begin{abstract}

We investigate the hypothesis that quasars formed together with the stellar 
populations of early-type galaxies. This hypothesis, in conjunction with
the stellar ages of early-type galaxies from population synthesis models,
the relation of black hole mass to bulge velocity dispersion,
and the velocity dispersion distribution of spheroids from the Sloan Digital Sky Survey,
completely determines the cosmic accretion history of supermassive black holes
and the redshift evolution of the characteristic luminosity.
On the other hand the precise shape of the luminosity function of quasars 
depends on the light curve of quasars and, in the optical, but no so much in X-rays, on
the covering factor of the dust surrounding the active nucleus. 
We find a plausible set of assumptions for which the coeval formation of 
supermassive black holes and elliptical galaxies is in good
agreement with the observed B-band and X-ray luminosity functions of quasars.

\end{abstract}

\section{Introduction}  
 
Several pieces of evidence point to a link between quasars and the
formation histories of elliptical galaxies: i) Spectrophotometric
studies of the host galaxies of optical quasars show isophotal
profiles consistent with the ${r^{1/4}}$ law and colours consistent
with red old stellar populations (McLure et al. 1999; Nolan et
al. 2001).  ii) A comparable number of active galactic nuclei are
embedded in dusty, violently interacting, starbursting systems, which
are thought to be the progenitors of elliptical galaxies.  iii) The
comoving density of bright quasars drops dramatically after $z\sim 2$.
By that time the stellar populations of bright ellipticals have been formed.
In contrast, star formation in discs extends to much lower
redshifts.  iv) There is mounting evidence for ubiquitous supermassive
black holes (SMBHs) in the nuclei of spheroids with a tight correlation of the
form $M_\bullet\propto\sigma^\eta$
between the BH mass $M_\bullet$ and the velocity dispersion of 
the host galaxy $\sigma$ (Ferrarese \& Merritt 2000;
Gebhardt et al. 2000a,b; Merritt \& Ferrarese 2001a,b; Tremaine et al. 2002).
BHs found in spiral galaxies are much less massive 
(Salucci et  al. 2000).

The existence of a link between quasars and elliptical galaxies is thus 
difficult to dispute.
However, the fuelling mechanism and the timing of the active phase are 
still not well understood. 
In the hierarchical picture for galaxy formation it is natural to assume that
the active phase coincides with the mergers responsible for determining 
the elliptical morphologies 
(Cattaneo, Haehnelt \& Rees 1999; Kauffmann \& Haehnelt 2000;
Haehnelt \& Kauffmann 2000; Cattaneo 2001; Kauffmann \& Haehnelt 2002; Volonteri, Haardt \& Madau 2002).
The active phase may not coincide with the epoch of star formation if 
much of the gas mass has already formed stars before the mergers.
Kauffmann \& Haehnelt succeeded in reproducing 
a good agreement with the luminosity function of quasars over the magnitude 
range $-28<M_B<-25$.
They were also able to reproduce the qualitative evolution of the 
luminosity function with redshift, but they could not account for
the observed sharp decline in the comoving density of quasars at $z\lsim 2$ 
(Fig.~9 of Kauffmann \& Haehnelt 2000).
In the hierarchical model the cooling of pristine gas forms 
larger and larger galaxies at low redshift. That
partly compensates the decrease in the merging rate and the consumption of
gas by star formation,
and has been indicated as the most plausible cause for this discrepancy.

Granato et al. (2001) explored the alternative assumption
that quasars were formed monolithically together with, or just after, the 
stellar populations of elliptical galaxies. They started from the observed 
luminosity function of quasars and used the assumption of the joint 
formation of quasars and spheroids to derive the rate of
formation of spheroids at high redshift. The predicted contribution of 
spheroids to SCUBA counts was found to be in agreement with the existing data.
The Granato et al. model is phenomenological because of the lack of
a model to incorporate monolithic collapse into cold dark matter cosmologies.

This paper continues the approach of Granato et al.
Our main aim is to  
determine if the assumption that quasars were formed together with 
the stellar populations of elliptical galaxies is consistent with the 
observed luminosity function of optical quasars.
The ingredients of the paper are:  
i) the relation of BH mass to bulge velocity dispersion,
ii) the velocity dispersion distribution of spheroids, and 
iii) the stellar ages of ellipticals inferred from population synthesis models.
In Section 2 we combine these ingredients to construct our model of the luminosity
function of quasars. In Section 3 we compare our model luminosity function with the
pure luminosity evolution model of Boyle et al. (2000) and the X-ray data of Cowie 
et al. (2003).

As a final point before we start, it is important to clarify that
we do not think of monolithic collapse as an alternative 
to hierarchical structure formation.
When we write that massive ellipticals were formed monolithically at high
redshift, we just mean that they have been passively evolving since 
$z\sim 2$ (e.g. Bernardi et al. 2003b,c,d).
This is not in conflict with the merging scenario if the mergers in 
which the stellar populations of these galaxies were formed happened 
at $z\gsim 2$.
Similarly, when we write that quasars were formed together with the stellar
populations of ellipticals, we do not mean the processes of BH accretion
and star formation are synchronous.
We just mean that if there is a delay of one with respect to the other,
this delay is short in comparison with the age of the Universe.
However, the delay can be significant in relation to the star formation
time-scale or the quasar life-time, and that would have important
consequences if we tried to predict the colours and infrared/sub-millimetre
properties of quasar hosts (e.g. Archibald et al. 2002).
See also Romano et al. (2002) and the new observations by Dietrich et al. (2003)
and Baldwin et al. (2003) for a discussion of the quasar timing in relation
to the star formation history of the host galaxy.

\section{A monolithic collapse model of the luminosity function of quasars}

\subsection{Ingredients}

In this section we shall develop a monolithic collapse model for 
the observed luminosity function of quasars and its evolution. 

In our model quasars have a distribution of masses 
$M_\bullet$ and times of formation $t_f$
(time means lookback time throughout this paper).
Quasars do not shine before their time of formation. 
At $t<t_{\rm f}$ quasars shine with 
\begin{equation}
\label{lightcurve}
L_{\rm bol}(t)=L_{\rm bol}(M_\bullet,t_{\rm f},|t-t_{f}|).
\end{equation}
The luminosity function at $t$ is the sum of the 
contributions from all quasars of all masses which formed at earlier 
times than $t$.  
Therefore, two elements combine to determine the luminosity function of quasars in our model:
i) the joint distribution of $M_\bullet$ and $t_{\rm f}$,
and ii) the quasar light curve (Eq.~\ref{lightcurve}).

The goal of this paper is to test the hypothesis that 
the distribution of $M_\bullet$ and $t_{\rm f}$ is determined by the distribution
of velocity dispersions and stellar ages of elliptical galaxies.
We note that this hypothesis only enters point (i).
Instead, the two physical quantities that determine the light curve are the life-time
and the radiative efficiency of quasars.
Our strategy is therefore as follows. Firstly, we shall use the early-type galaxy
sample in the Sloan Digital Sky Survey (SDSS, Bernardi et al. 2003a) to infer a joint distribution 
for $M_\bullet$ and $t_{\rm f}$ (Section 2.2).
Then we determine what light curve gives a good fit
to the observed blue luminosity function when taken in combination with this distribution.
The idea is that if we find a good fit by using a quasar life-time and a
radiative efficiency which are consistent with those favoured by the literature, 
then we have an argument for believing that we have a tenable model for the distribution of 
masses and formation times of SMBHs.

\subsection{Black hole formation times and masses from the SDSS}

Our model relates the masses and formation times of quasars to those 
of early-type galaxies.  We discuss the distribution of quasar masses 
first, and then discuss the distribution of formation times at fixed 
mass.  

BHs of mass $M_\bullet$ are thought to form in galaxies 
with velocity dispersion $\sigma$ given by 
\begin{equation}
\label{trema}
M_\bullet=1.7\times 10^8h_{65}^{-1}\sigma_{200}^5M_\odot,
\end{equation}
($\sigma_{200}\equiv\sigma/200{\rm\,km\,s}^{-1}$).
Therefore, if the distribution of stellar velocity dispersions is 
known, the distribution of BH masses can be 
obtained by a change of variables.  
The power in Eq.~(\ref{trema}) is still the object of a controversy.
We have used a power of 5 because it gives the best fit to the luminosity function of
quasars (see the next Section), but a power of 4 would still give an acceptable fit to
the luminosity function, specially if one readjust the stellar ages within their large
margins of error.

Bernardi et al. (2003a) present a sample of nearly 9000 early-type galaxies 
at redshifts $z<0.3$, selected from the SDSS by using 
morphological and spectral criteria. In a series of papers they study the 
properties of early-type galaxies such as the luminosity function, and 
various early-type galaxy correlations in multiple bands 
(Bernardi et al. 2003b), the Fundamental Plane
and its dependence on wavelength, redshift and environment 
(Bernardi et al. 2003c), the colours, and how the chemical 
composition of the early-type galaxy population depends on redshift 
and environment (Bernardi et al. 2003d).
In this SDSS sample, the distribution of the velocity dispersion $\sigma$ is 
$$N(\sigma){\rm\,d}\sigma=1.388\times 10^{-3}h_{65}^3\left({\sigma\over94.4{\rm km\,s}^{-1}}\right)^{6.3}\times$$
\begin{equation}
\label{sigmadist}
\times{\rm\,exp}\left[-\left({\sigma\over 94.4{\rm km\,s}^{-1}}\right)^{2}\right]{{\rm d}\sigma\over\sigma}
\end{equation}
(Sheth et al. 2003).
By inserting the $M_\bullet(\sigma)$ variable transformation 
into the $N(\sigma)$ from the SDSS we have a model for the distribution 
of $M_\bullet$.  

The next step is to determine the distribution of formation times.  
We assume that this is given by the distribution of formation times
for the stellar populations of early-type galaxies. 
As we describe below, this too can be determined from the SDSS sample.  

The signal-to-noise ratios of the spectra in the SDSS sample are 
substantially smaller than the $S/N \sim 100$ required to estimate the 
Lick indices reliably. One of the results of Bernardi et al. (2003d) is a 
library of 182 co-added spectra, which contains spectra that represent 
a wide range of early-type galaxies.  
Specifically, galaxies were divided into six bins for each of the following 
quantities/characteristics:
luminosity, velocity dispersion, size, redshift and environment.  
The spectra in each bin were co-added (this was a straight average, 
i.e., no weighting by S/N was used).   The bin sizes were chosen 
to be small enough that each composite spectrum is truly representative 
of galaxies with similar properties, but large enough to ensure that 
the resulting co-added spectra have $S/N>50$ (the mean $S/N$ per pixel
is 129 with a rms scatter of 37, and the maximum $S/N$ is 238).   
The co-added spectra were used to obtain reliable estimates of 
absorption-line strengths and to investigate the chemical evolution 
and star formation histories of early-type galaxies.

By comparing the measured line indices (Mg$b$ and H$_\beta$) 
reported in Table~4 of Bernardi et al. (2003d) with the single 
burst stellar population models of Worthey et al. (1994)
it is possible (by linear interpolation) to determine the age and the 
metallicity of each spectrum.  
There are at least two caveats associated with these age estimates.  
First, the age estimates make use of models which assume that the bulk 
of the stellar population formed in a single burst.  
Second, the models assume solar values for the ratio of $\alpha$ element 
abundances relative to iron, whereas early type galaxies have 
[$\alpha$/Fe] enhanced relative to the solar value.  
Bernardi et al. (2003d) describe a simple correction for this, but 
their correction is by no means secure.   
However, by applying that simple correction the resulting galaxy age 
estimates given by the (Mg$b$ and H$_\beta$) diagram are consistent 
with the ages obtained by using the (Fe and H$_\beta$) measurements.
In addition, the resulting age estimates from galaxies in different 
redshift bins are consistent with one another, once the cosmic time
difference between different redshifts has been accounted for.
That suggests 
that these single-burst ages offer at least a self-consistent picture 
of the passive evolution of the early-type galaxy population.  These 
age estimates are also consistent with the evolution of the luminosity 
function (Bernardi et al. 2003b), the colours (Bernardi et al. 2003d) 
and the Fundamental Plane (Bernardi et al. 2003c).  
It is these single-burst ages which we use in what follows.  

Table~4 of Bernardi et al. (2003d) also provides the velocity 
dispersion, $\sigma$, and the redshift of the spectra. 
Thus, for each redshift bin, we can
compute the mean age as function of velocity dispersion. The result
is shown in Fig.~1 (open squares show the measurements from spectra at 
$z < 0.075$ and filled circles refer to spectra in the redshift interval 
$0.075 < z < 0.1$). The error bars show the rms scatter divided by $\sqrt{N}$. 
Notice that there is a strong correlation between stellar velocity 
dispersion and age:  the galaxies with the highest values of $\sigma$ 
contain the oldest stellar populations.
Other analyses of the stellar populations of early-type galaxies 
come to a similar conclusion.  E.g., Thomas, Maraston \& Bender (2002) 
find a similar trend, although the actual ages they report are slightly 
older.  If we use their models to estimate ages in the SDSS data, then 
we also find higher formation redshifts, and a shallower relation 
between age and $\sigma$.  

\begin{figure} 
\centerline{\psfig{figure=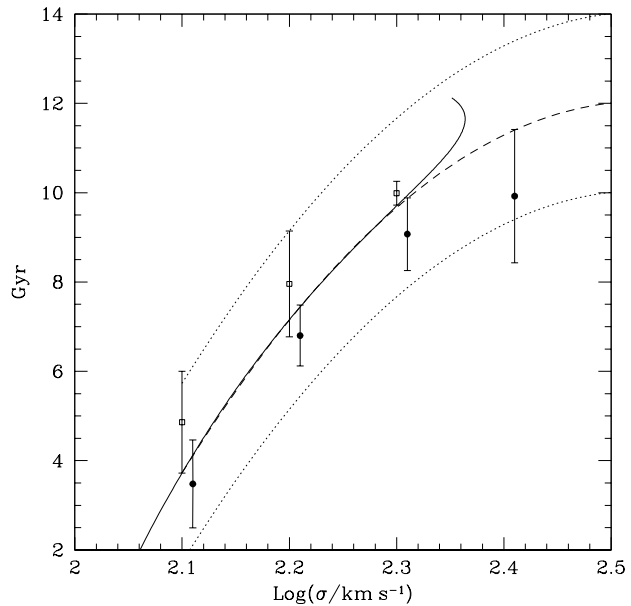,width=0.48\textwidth,angle=0}}
\caption{\small The stellar age - velocity dispersion dispersion relation in 
the galaxy sample of Bernardi et al. (2003d; open squares: $z<0.075$; 
filled circles: $0.075<z<0.1$). The solid line shows the expected mean 
age - $\sigma$ relation if the stars and the supermassive black holes of 
elliptical galaxies were formed at the same cosmic time (Eqs.~2 and 9, see Section 3).
The dashed line (Eq.~\ref{fitage}) is a fit to the stellar ages of ellipticals
(shown by the points with error bars).
This fit is used to calculate the optical luminosity function (Fig.~2) and 
the cosmic accretion rate (Fig.~4) of quasars. The dotted lines 
correspond to a spread of $\pm\,2\,$Gyr about the dashed line.}
\end{figure}

The dashed line in Fig.~1 shows a simple fit to the 
relation between age and $\sigma$:
\begin{equation}
\label{fitage}
\bar{t}=\left[{12.05-45\left({\rm Log} {\sigma\over 339{\rm\,km\,s}^{-1}}\right)^2}\right]{\rm\,Gyr}.
\end{equation}
Fig.~1 also shows that there is a spread in ages at 
fixed velocity dispersion of $\Delta t\sim 2$~Gyr, which is 
approximately independent of $\sigma$ (the dotted lines show 
$\bar{t}\pm\Delta t$).  We assume that this spread is Gaussian.  
Thus, the SDSS sample provides us with a measurement of the joint 
distribution of early-type galaxy velocity dispersions and formation 
times.  Therefore, we now have a model for the joint distribution 
of BH masses and formation times which we can use to construct 
our model of the quasar luminosity function.

\subsection{The quasar light curve}

We assume that a quasar of mass $M_\bullet$ which forms at $t_{\rm f}$ shines 
at $t\le t_{\rm f}$ with luminosity:
\begin{equation}
\label{lightcurve}
L_{\rm bol}=\dot{m}_0 L_{\rm Edd}(M_\bullet)
{\rm exp}\left(- {|t-t_{\rm f}|\over t_{\rm QSO}} \right), 
\end{equation}
where $\dot{m}_0 < 1$ is a free parameter, which may depend on 
$M_\bullet$ and $z$. This is our way of specifying Eq.~(1).
Here
\begin{equation}
t_{\rm QSO}=0.44{\epsilon\over\dot{m}_0}{\rm\,Gyr}
\end{equation} 
is the characteristic life-time of quasars determined by the condition
$\int L_{\rm bol}{\rm\,d}t=\epsilon M_\bullet{\rm c}^2$, where ${\rm c}$
is the speed of light and $\epsilon\sim 0.1$ is the radiative efficiency of quasars.

In reality the quasar light curve will not be a simple function such that in Eq.~(\ref{lightcurve}).
Very probably there will be two phases: a rising part, in which the luminosity is limited by rate 
at which the BH can grow (the Eddington limit), and a fading part, in which the luminosity drops as
the quasar runs out of fuel. Here we do not follow the mass growth of the BH, so we do not have the
mass as a function of time. Instead we take a model light curve $L_{\rm bol}(t)$ such as that given by
Eq.~(\ref{lightcurve}), which contains a number of free parameters including the final BH mass $M_\bullet$. 
The careful reader will have noticed that with this modelling $L_{\rm bol}$ can never be equal to 
$L_{\rm Edd}(M_\bullet)$, if the Eddington limit is to be respected,
because a quasar cannot have already reached its final mass at the peak of its 
activity. We chose this model because $L_{\rm Edd}(M_\bullet)$ is the best estimate of the order of magnitude
of the maximum luminosity a quasar can ever had. Any constant of proportionality will be reabsorbed in the value
of $\dot{m}_0$

The conversion of bolometric luminosities into blue magnitudes and X-ray luminosities
is performed by assuming the spectral energy distribution in Elvis et al. (1994), for which 
a bolometric luminosity of $L_{\rm bol}=1.3\times 10^{46}{\rm erg\,s}^{-1}$
corresponds to a blue magnitude of $M_B=-25$ and a total luminosity in the 2-8\,keV band
of $L_X=4\times 10^{44}{\rm erg\,s}^{-1}$.
Also notice that this is the emission corresponding to a quasar of $M_\bullet=10^8M_\odot$
which is radiating at the Eddington limit.

In Section 2.5 we shall tune the free parameters $\epsilon$ and $\dot m_0$ to fit the 
observed blue luminosity function,
given the joint distribution of $M_\bullet$ and $t_f$ determined in Section 2.2.

\subsection{Luminosity-dependent obscuration}

There is one final subtlety that we must incorporate into our 
model of the quasar luminosity function.  

The development of spectropolarimetry imaging (see Antonucci 2002), the 
discovery of type 2 quasars in X-ray surveys 
(e.g. Norman et al. 2002) and the infrared spectroscopy of ultra-luminous infrared galaxies 
(see Lutz 2000; Sturm et al. 1997) and radio galaxies (Hill, Goodrich \& De Poy 1996) have shown that many quasars
do not contribute 
to optical counts because of the presence of various sorts of absorbers along the line of sight:
dusty tori, narrow-emission-line clouds, clouds of gas in the host galaxy, etc. 
(see Rowan-Robinson 1995). 
Scheuer (1987) and Barthel (1989) used the dusty torus model to propose that radio-loud quasars 
and radio galaxies are the same type of object -- the former observed pole-on, the latter with 
the torus's axis close to the plane of the sky. In this way, they successfully extended to radio sources
the orientation model
introduced by Antonucci \& Miller (1985) to unify Seyferts 1 and Seyferts 2.

There is evidence that the ratio of obscured to unobscured quasars is a function of the 
bolometric luminosity. E.g. for radio-loud objects Lawrence (1991) found that at the
highest radio luminosities the numbers of broad-line and narrow-line quasars are about even,
while the ratio of narrow-line and reddened broad-line quasars is substantially higher at lower
radio powers. His explanation was that, as the quasar grows more and more 
powerful, it heats the surrounding dust to the point of sublimation. 
Consequently, the dusty torus is eroded from within and opens up its angle.
This is called the receding torus model and predicts that the tangent of the torus's opening
angle scales as a power of the radio luminosity.

While the absorbing material may not always have the form or the geometry of a dusty torus,
the finding that the fraction of broad-line quasars is much lower at lower source power is
strongly supported by recent X-ray studies.
An extremely deep ($\sim 1\,$Ms) CHANDRA X-ray study of the Hubble Deep Field North and its 
environs has resolved 90\% of the X-ray background from the targeted sky area 
into 370 discrete X-ray sources (Brandt et al. 2001). 
Barger et al. (2002) have been able to collect good quality optical spectra, and therefore
spectroscopic redshifts, for 170 of these sources.
It is obvious from this work and from another more recent paper of the same group (Cowie et al. 2003)
that broad emission lines are much more common in objects that are very luminous in hard X-rays.
The majority of objects with low hard X-ray luminosity show absorption features at soft
X-rays consistent with very large column densities ($N_H>10^{23}{\rm\,cm}^{-2}$) as well as 
very steep (highly absorbed) infrared/optical spectra when compared with the median spectral energy
distribution of Elvis et a.(1994). Also see Mainieri et al. (2002) for the relation between
X-ray absorption and optical reddening/obscuration.

In our Monte Carlo simulations, we assume that the probability of a quasar being observed is
proportional to the opening cone of a torus, the opening angle $\theta$ of which is
determined by the relation
\begin{equation}
{\rm tan}\,\theta=({ L_{\rm bol}/L_{\pi/4}})^\eta,
\end{equation}
independently of whether that is the real geometry of the obscuring screen.
Here we treat $L_{\pi/4}$ and $\eta$ as two additional free parameters, 
besides $\epsilon$ and $\dot{m}_0$, which we use to fit the observed blue luminosity function of
quasars. In a forthcoming publication Cattaneo et al. (2003) will analyse the fraction of
obscured quasars as a function of $L_{\rm bol}$ for X-ray selected quasars.

\section{Comparison with the observed optical and X-ray luminosity functions}

Following Boyle, Shanks \& Peterson (1988),
it is usual to fit the blue luminosity function of quasars to a double power law 
function of the form:
\begin{equation}
\label{ple}
\phi(M_B,z)={\phi_*\over
10^{0.4[M_B-M_B(z)](\alpha+1)}+10^{0.4[M_B-M_B(z)](\beta+1)}}
\end{equation}
where the only dependence on redshift is in 
the characteristic magnitude $M_B=M_B(z)$ at which the exponent of the power changes
(pure luminosity evolution).

Boyle et al. (2000) have analysed over 6000 quasars from the 2dF QSO Redshift Survey.
For a cosmology with $\Omega=0.3$ and $\Lambda=0.7$ (assumed throughout this paper)
their best fit parameters are: 
$\phi_*=2.9\times 10^{-6}h_{65}^5{\rm\,Mpc^{-3}mag^{-1}}$,
$\alpha=-1.58$, $\beta=-3.41$, and
\begin{equation}
\label{ple2}
M_B(z)=-22.08-2.5(1.36z-0.27z^2)+5{\rm Log}h_{65}.
\end{equation}
Here $h_{65}\equiv H_0/65{\rm\,km\,s^{-1}\,Mpc^{-1}}$.
The lines in Fig.~2 show the pure evolution model of Boyle et al. from $z=2.3$ (upper solid
line) to $z=0.61$ (lower dashed line). 

Our best fit to the pure luminosity evolution model of
Boyle et al. (2000) is shown by the symbols in Fig.~2 (open squares, filled circles and hexagons)
and is obtained for a BH radiative efficiency of $\epsilon=0.1$ and a power of $\eta=0.88$
in the scaling of the torus opening angle with luminosity (Eq.~7).
To obtain the best fit  
we normalise the relation between opening angle and luminosity,
which means we set the value of $L_{\pi/4}$,
so that a quasar with $M_B=-24$ has a 50\% chance of being observed.
We find no need to change these three quantities ($\epsilon$, $\eta$,  $L_{\pi/4}$) with redshift.
Instead, to find a good fit, we need to allow for a dependence on redshift in the value of
$\dot{m}_0$.
We find a good agreement with the luminosity function of Boyle et al.
if low redshift quasars begin their growth at 
a lower fraction of the Eddington limit, or, put in other words, 
if the life-time of quasars is longer at low redshift.
The best agreement with the observations is for 
$\dot{m}_0=0.7$ at $z\sim 2$, $\dot{m}_0=0.4$ at $z\sim 1.3$, 
and $\dot{m}_0=0.2$ at $z\sim 0.7-0.8$. We shall comment on  this finding in Section 5.
We only fit the luminosity function at $z<2.3$ because using 
the model at higher redshifts is equivalent to extrapolating 
the fit provided by Eq.~(\ref{fitage}) to $\sigma>250{\rm km\,s}^{-1}$ 
(see Fig.~1), and it is not clear that this extrapolation is 
warranted.

\begin{figure} 
\centerline{\psfig{figure=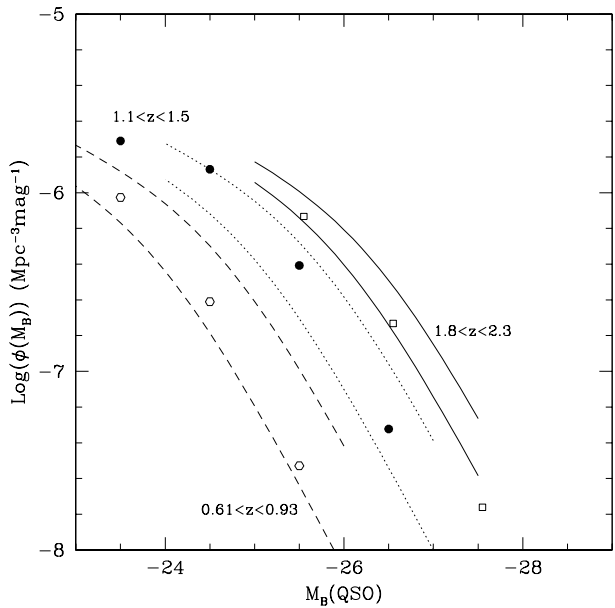,width=0.48\textwidth,angle=0}}
\caption{\small Simulated blue luminosity functions of quasars: $0.61<z<0.93$ (open hexagons);
$1.1<z<1.5$ (filled circles); $1.8<z<2.3$ (open squares).
The lines are shown for comparison and give the pure luminosity evolution phenomenological model of 
Boyle et al. (2000; Eqs.~\ref{ple}-\ref{ple2}).}
\end{figure}

As a check, we show that the evolution of the characteristic blue luminosity in the Boyle et al.
model is directly consistent with the $\sigma$-age relation for early-type galaxies.
For this purpose, let us call $L_{\rm bol}(z)$ 
the characteristic luminosity as a function of redshift, 
which we infer from Eq.~(\ref{ple2}) by using
the bolometric corrections in Elvis et al. (1994). 
We use the Eddington limit to associate a characteristic BH mass $M_\bullet$ to a 
characteristic luminosity $L_{\rm bol}$.
Similarly, we use our model cosmology ($\Omega=0.3$ and $\Lambda=0.7$, $h_{65}=1$)
to calculate the lookback time $t$ of an event at redshift $z$.
So, we have determined $M_\bullet(t)$, the characteristic mass of quasars that form at 
lookback time $t$.
With this double transformation and by using Eq.~(2), which relates the 
BH mass to the velocity dispersion, the function 
$M_B(z)$ of Eq.~(\ref{ple2}) 
becomes the function $\sigma(t)$ displayed as the solid line in Fig.~1.  
This figure shows consistency between the $\sigma$-age relation of 
early-type galaxies derived from the stellar population synthesis models and the cosmic
evolution of the characteristic luminosity of quasars.
The solid line turns back at ${\rm Log}(\sigma/{\rm km\,s}^{-1})\simeq 2.36$
because $M_B(z)$ reaches a minimum at $z\simeq 2.52$.
The point is that the blue luminosity function of bright quasars rises up to $z\sim 2.5$
and then drops.
In our model we interpret that as consequence of the fact that $\bar{t}(\sigma)$ reaches 
a maximum at $\bar{t}_{\rm max}\sim 12\,$Gyr because it is difficult for a galaxy with  
${\rm Log}(\sigma/{\rm km\,s}^{-1})\gsim 2.5$ to be assembled in less than $\sim 2.5\,$Gyr
(14.46\,Gyr is the age of the Universe in the cosmology used for this paper).
Only a Gaussian tail of increasingly rarer galaxies form at $t>\bar{t}_{\rm max}$.
So in our model the luminosity function drops at $z\gsim 2.5$ because
the number of quasars, and not their characteristic luminosity, drops.
However, in the Boyle et al. s pure luminosity evolution model a fall in the characteristic 
luminosity is the only way to reproduce the decrease of the blue luminosity function.
Consequently, we do not think of the discrepancy at ${\rm Log}(\sigma/{\rm km\,s}^{-1})>2.36$
as a failure of our model. 
Moreover, we remark that in our model a large fraction of the quasars at lookback time
$t\sim 11.5\,$Gyr are in galaxies with ${\rm Log}(\sigma/{\rm km\,s}^{-1})\sim 2.3$
and not ${\rm Log}(\sigma/{\rm km\,s}^{-1})\sim 2.4$.
That is because the former are much more numerous than the latter (Eq.~\ref{sigmadist}) and 
because we have assumed that there is a large spread ($\sim 2\,$Gyr) around $\bar{t}$
(see comment at the end of this section).
It can be argued that the test performed in this paragraph 
is redundant because if our model reproduces the
luminosity function and its redshift evolution, then automatically it also reproduces the redshift
evolution of the characteristic magnitude at which the slope of the luminosity function changes.
However, the interesting point of this check is that the comparison in Fig.~1 is independent of any
free parameter. So the comparison in Fig.~1 probes the distribution of BH masses and
formation times directly, independently of the details of the light curve (as long as $\dot{m}_0$
is not much smaller than one).

A comparison of Figs.~2 and~3 show that while Eq.~(\ref{fitage}) is the driving factor in determining
the evolution of the comoving density of bright quasars, the evolution of faint sources
is almost entirely driven by the assumption made about the obscuration.
 
\begin{figure} 
\centerline{\psfig{figure=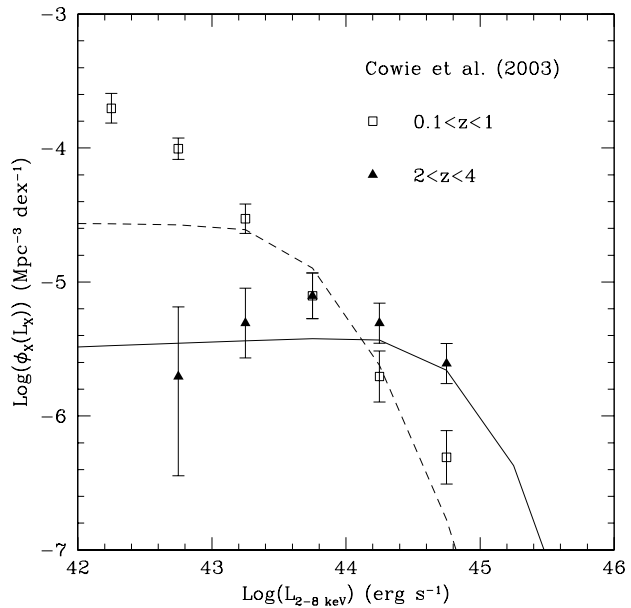,width=0.48\textwidth,angle=0}}
\caption{\small Simulated X-ray luminosity function of quasars at $0.1<z<1$ (dashed line) and $2<z<4$
(solid line) compared with the data points by Cowie et al. (2003).}
\end{figure}
Now all parameters are fixed and we no longer have any freedom for adjusting
the X-ray luminosity function.
With the bolometric correction for the X-band inferred 
from the Elvis et al. (1994) spectral energy distribution and 
by assuming that we can neglect X-ray absorption,
we find the X-ray luminosity function shown by the lines in Fig.~3,
which we compare with the data points of Cowie et al. (2003) from the CHANDRA Deep
Field North (CDFN).
The difference between the observed and simulated $0.1<z<1$ luminosity functions at 
$L_X<10^{43}{\rm\,erg\,s}^{-1}$ is not significant because that part of the diagram receives a
contribution from active galactic nuclei in the bulges of spirals and Eq.~(\ref{sigmadist}) only contain 
information about early-type galaxies.

As a final comment, we note that, in order to reproduce the quasar luminosity function successfully,
it is necessary that a large part of the age spread in Eq.~(\ref{fitage}) is intrinsic, and not just 
the consequence of measurement errors. Otherwise, the only BHs that would be forming
at lookback time $t$ would be those with $M_\bullet\le M_\bullet[\sigma(t)]$, where $M_\bullet(\sigma)$
and $\sigma(t)$ are given by Eqs.~(\ref{trema}) and (\ref{fitage}).
So there would be a sharp cut-off in the luminosity function of quasars at $L_{\rm Edd}[M_\bullet(t)]$,
which is not found in the observation (in our model the life-time of quasars is so short that the
contribution to the luminosity function at lookback time $t$ from quasars formed at higher redshifts
is negligible).

\section{Cosmic accretion history of supermassive black holes}

The previous section showed that a picture in which the formation 
of SMBHs is closely tied to that of elliptical 
galaxies, is consistent with the observed quasar luminosity function.  
We now explore what this implies for the cosmic accretion 
history of SMBHs.  

The BH mass per comoving volume in galaxies with velocity dispersion 
between $\sigma$ and $\sigma+{\rm d}\sigma$ is 
$M_\bullet N(\sigma){\rm\,d}\sigma$, where $N(\sigma){\rm\,d}\sigma$ 
is given by Eq.~(3) of the previous section.  
In the previous section we modelled the formation time distribution 
of elliptical galaxies as a Gaussian with mean $\bar{t}(\sigma)$ and 
standard deviation $\Delta t(\sigma)\sim 2$~Gyr inferred from the SDSS 
measurements (Eq.~4, dashed line in Fig.~1).  
Therefore, in this model, the cosmic accretion rate of SMBHs is:
\begin{equation} 
\label{carr}
\dot{\rho}_\bullet(t)=
\int{M_\bullet(\sigma)N(\sigma)\over\sqrt{2\pi}\Delta t(\sigma)}
{\rm exp}\left\{-{[t-\bar{t}(\sigma)]^2\over 2\Delta t^2(\sigma)}\right\}{\rm\,d}\sigma.
\end{equation}

This estimate of the cosmic accretion rate, which is based on 
measurements of the properties of early-type galaxies, can be 
compared with a lower limit which comes directly from the quasar 
luminosity function:
\begin{equation}
\label{sct}
{{\rm d}\rho_\bullet\over{\rm d}z}     
={1\over\epsilon{\rm c}^2}
\int L_{\rm bol}(M_B)\phi(M_B,z){\rm\,d}M_B
\end{equation}
where $L_{\rm bol}(M_B)$ is the bolometric luminosity corresponding to a 
quasar of blue magnitude $M_B$, ${\rm c}$ is the speed of light, $\epsilon$
is the radiative efficiency of quasars, and $\phi(M_B,z)$ is the luminosity 
function of quasars at redshift $z$ (So{\l}tan 1982; Chokshi \& Turner 1992).
With the luminosity function of Boyle et al. (2000; Eqs.~\ref{ple}-\ref{ple2}),
Eq.~(\ref{sct}) becomes:
\begin{equation}
\label{sct2}
\dot{\rho}_\bullet
={1\over\epsilon{\rm c}^2}(1.08\phi_*)L_{\rm bol}[M_B(z)]
\int {{\rm d}\zeta\over \zeta^{\alpha-1}+\zeta^{\beta-1}},
\end{equation}
where the factor of 1.08 originates from converting from magnitudes into luminosities.

Fig.~4 compares the cosmic accretion rates inferred from Eqs.~(\ref{carr}) and (\ref{sct2}).
Our model predicts that optical quasars are not a good a tracer of the cosmic accretion
history of SMBHs. 
Our model does reproduce the observed peak of the characteristic luminosity at $z\sim 2.5$ (Fig.~2),
but, as seen in Fig.~4,
the largest contribution to the cosmic density of SMBHs does not come from the very bright
quasars at $z\sim 2.5$. It rather comes from intermediate luminosity quasars at $z\sim 1-1.5$,
which are much more numerous, although optical counts underestimate their number, since they
are much more frequently obscured than their $M_B=-26$ counterparts.
This result is robust in the sense that it is independent of $\dot{m}$, $L_{\pi/4}$ and $\eta$,
and continues to hold for reasonable changes of the fitting function (\ref{fitage}).
We shall expand this point in the next section.

\begin{figure} 
\centerline{\psfig{figure=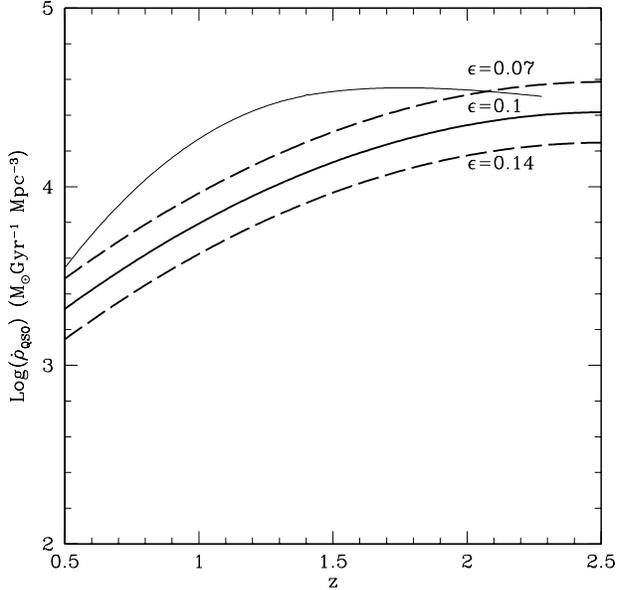,width=0.48\textwidth,angle=0}}
\caption{\small The thick lines show the cosmic accretion rate inferred from 
Eq.~(\ref{sct2}) for different values of the radiative efficiency.
The thin line shows the cosmic accretion rate from Eq.~(\ref{carr}).}
\end{figure}

\section{Discussion and conclusion} 

In this paper we have demonstrated that a very simple model in which quasars 
are short lived and form monolithically with the stellar populations of 
elliptical galaxies reproduces an excellent agreement with the optical and X-ray
luminosity functions of quasars if two very plausible and motivated 
assumptions are made: i) the peak luminosity is $\sim L_{\rm Edd}$
at $z\sim 2.5$ but drops to $\sim 0.1L_{\rm Edd}$ at $z\sim 0$, and 
ii) the less powerful quasars are more likely to
be absorbed at optical wavelengths by large column densities on the line of sight. 
Haiman \& Menou (2000) and Kauffmann \& Haehnelt (2000) also
found (from a completely different approach) that one requires longer 
accretion time scales at low redshift to reproduce the 
optical luminosity function of quasars (assumption i). 
That makes sense because cosmic structures that form at high redshift are denser
and therefore have shorter dynamical times.
However, (ii) is the key assumption that 
has allowed to find a good fit to the blue luminosity function of quasars.  
This hypothesis can be checked by assessing the fraction of objects with
broad emission lines at different values of the intrinsic  power. 
To construct a sample of objects with the same intrinsic power one has to 
find a band in which one can see the unobscured quasar emission. 
Studies in this sense have done by using: i) Paschen spectroscopy 
(Hill, Goodrich \& De Poy 1996): broad lines in the Paschen series should 
appear through the dust even for extinction as high as $A_V=15$,
ii) radio surveys (for radio-loud quasars; e.g. Simpson 1998), 
and iii) X-ray surveys (e.g. Barger et al. 2002).
These studies confirm that the less powerful active galactic nuclei are 
less likely to display detectable broad line emission,
but it is important to obtain a more quantitative statement to be able to
remove some of the uncertainty, which is always present in models with a number
of free parameters. 
Moreover, we need 
to understand whether not only the column density but also 
the type of the absorber depends on the range of power or changes 
with redshift.
However, it should be remarked that, while we need to make a number of assumptions
about the light curve in order to produce a model for the quasar luminosity 
function, it is also true that there are some results which are to a large extent 
independent of the parameters and the parameterisation (Figs. 1 and 4).
They are the results that originate directly from our main assumption that SMBHs
were formed at the same epoch of the stellar populations of ellipticals.

\begin{figure} 
\label{clustering} 
\centerline{\psfig{figure=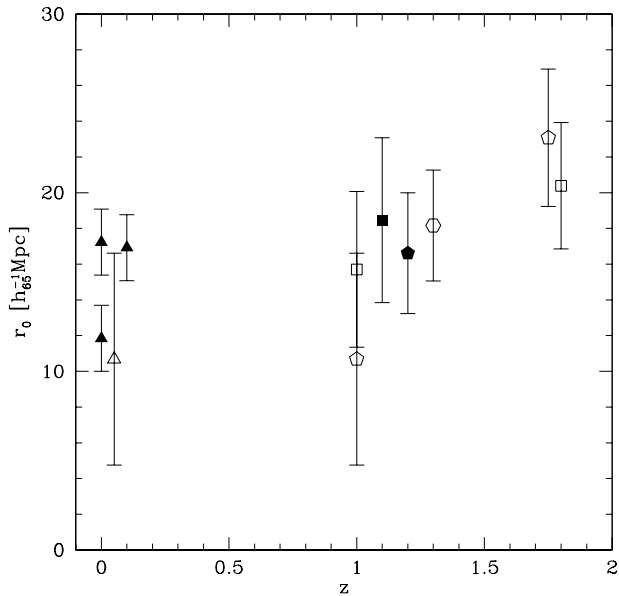,width=0.48\textwidth,angle=0}}
\caption{\small The clustering length for quasars with $M_B<-23$
(open points; La Franca, Andreani \& Cristiani 1998) compared with the 
clustering length of local ellipticals and $z\sim 1$ extremely red 
objects (filled points; Daddi et al. 2001).}
\end{figure}

A key ingredient of our model is the $\sigma$-age empirical relation, which assigns older
ages to the most massive ellipticals.
At a superficial look that seems to contradict hierarchical structure formation.
Granato et al. (2001) proposed a theoretical explanation for this finding.
Stellar feedback can
prolong star formation in low mass haloes, while in massive haloes the 
potential well is much deeper, and it is therefore more difficult for the gas
to be reheated or ejected. So the less massive haloes, which are
statistically the first ones to virialise, do not form many stars.
Stars form much more rapidly and easily in the most massive haloes.
That produces an antihierarchical behaviour of the baryons,
where the most massive galaxies are the first ones to form stars. 
That antihierarchical behaviour is what makes the Granato et al. model close to a 
monolithic one.
Moreover Blanton et al. (1999, 2000) have argued that the formation of large systems 
is prohibited at low redshift in hot cluster gas.
When environmental effects of that sort are included,
inverted ages arise naturally in hierarchical model of structure formation. 

Here we have implicitly assumed that each SMBH forms in one short (0.06-0.2Gyr) episode.
Quasars cannot radiate for much longer than that without producing an excess of SMBHs
(we have checked ourselves that if quasars with $M_B<-26$ which are active at $z=2.5$
remained switched on for the rest of the life of the Universe, that would leave far too
large a number of relic BHs with $M_\bullet>2\times 10^{10}M_\odot$).
However, the accretion could still be spread over most of the life of the
Universe as a sequence of intermittent progressively less violent bursts. 
In fact, it is plausible that many low redshift quasars are triggered by the refuelling of 
SMBHs formed
at an earlier epoch.
That is particularly true for BHs that are found in rich environments (Cattaneo 2002).
There are, nevertheless, two points that should not escape our attention.
First of all,
the decrease of the quasar clustering length at low redshift is accompanied by a similar decrease
in the correlation length when we pass from $z\sim 1$ extremely red objects (EROs)
to local $z\sim 0$ ellipticals (Fig.~5).
There is evidence that the 
population of low redshift quasars 
is dominated by the formation of new lower mass BHs in shallower potential wells, which we identify with
galaxies with $\sigma\sim 110-130{\rm\,km\,s}^{-1}$.
Haehnelt, Natarajan \& Rees (1998), Haiman \& Loeb (1998), La Franca, Andreani \& Cristiani (1998),
Kauffmann \& Haehnelt (2000) and Haiman \& Hui (2001) reached a similar conclusion from Press-Schechter theory,
arguing for the continuous formation of new BHs and short duty-cycles.
Secondly, Shields et al. (2003) have used width measurements of quasar emission lines to probe the $M_\bullet(\sigma)$
relation
as a function of redshift (BH masses were derived from the continuum luminosity and the broad H$\beta$ emission
line of the quasar with a method known as reverberation mapping, whereas the velocity dispersions of the host galaxies
were determined from the width of the narrow [O III] lines. They found that the relation out to $z\sim 3$ is
indistinguishable from that found locally and concluded that this is consistent with the idea that SMBHs and
bulges grow simultaneously. Thus, if BHs form after the host bulges, then the delay has to be short 
in comparison with the Hubble time.

Perhaps the most interesting result of this paper is that optical quasars 
are biased tracers of the cosmic accretion rate of SMBHs 
(Fig.~4). Although the optical luminosity function peaks at $z\sim 2.5$ (Fig.~2), most 
of the accretion happens later 
($z\sim 1.3-1.5$, and even lower, when the contribution from Seyferts in Sa-Sb bulges is considered) 
in lower mass BHs. 
This accretion is not properly accounted for in optical studies
because it is highly obscured. 
The importance of obscured accretion is demonstrated by the fact that optical quasars account for
$\sim 20\%$ of the hard X-ray background (e.g. Salucci et al. 1999).
Our finding is also consistent with the 
recent discovery from ultra-deep CHANDRA and XMM counts that most of the 
X-ray background come from sources around $z\sim 1$ and not $z>2$ 
(Rosati et al. 2002; Hasinger 2002; Franceschini, Braito \& Fadda 2002),
and also from the X-ray selected quasar sample from the CDFN, where Cowie et al. (2003)
find that the most powerful quasars are much more likely to show broad emission lines.
Our confidence about this result also relies on the fact that we are able to reproduce
a reasonable agreement with the observed X-ray luminosity function of quasars (Fig.~3).
This partly explains why studies of 
quasars in the context of semi-analytic galaxy formation have had so many
problems finding the peak at the right redshift: it is not because the model assumptions
were wrong in themselves, but that the power dependence of the obscuration 
was not properly considered.

\section{Acknowledgements}

We acknowledge discussion with Claudia Maraston, Daniel Thomas and Pierluigi Monaco.
Moreover we thank Luigi Danese and Gian Luigi Granato for comments on the manuscript.
This research has been supported by the European Commission under the Fifth Framework Program.


\begin{thebibliography}{99}
\bibitem{}
Antonucci R., 2002, in J. Trujillo-Bueno, F. Moreno-Insertis, F. S\'anchez, eds.,
Proceedings of the XII Canary Islands Winter School of Astrophysics.
Cambridge University Press, Cambridge, p. 151
\bibitem{}
Antonucci R.RJ., Miller J.S., 1985, ApJ, 297, 621
\bibitem{} 
Archibald E.N., Dunlop J.S., Jimenez R., Fria\c{c}a A.C.S., McLure R.J., Hughes D.H.,
2002, MNRAS, 336, 353
\bibitem{}
Baldwin J.A., Hamann F., Korista K.T., Ferland G.J., Dietrich M., Warner C., 2003, ApJ, 583, 649
\bibitem{}  
Barger A.J., Cowie L.L., Brandt W.N., Capak P., Garmire G.P.,
Hornschemeier A.E., Steffen A.T., Wehner E.H., AJ, 124, 1839
\bibitem{}
Barthel P.D., 1989, ApJ, 336, 606
\bibitem{}
Bernardi M., Sheth R. K., Annis J., et al., 2003a, AJ, in press (astro-ph/0301631)
\bibitem{}
Bernardi M., Sheth R. K., Annis J., et al., 2003b, AJ, in press (astro-ph/0301624)
\bibitem{}
Bernardi M., Sheth R. K., Annis J., et al., 2003c, AJ, in press (astro-ph/0301626)
\bibitem{}
Bernardi M., Sheth R. K., Annis J., et al., 2003d, AJ, in press (astro-ph/0301629)
\bibitem{}
Blanton M., Cen R., Ostriker J.P., Strauss M.A., 1999, ApJ, 522, 590
\bibitem{}
Blanton M., Cen R., Ostriker J.P., Strauss M.A., Tegmark M., 2000, ApJ, 531, 1
\bibitem{}
Boyle B.J., Shanks T., Croom S.M., Smith R.J., Miller L., Loaring N.,
Heymans C., 2000, MNRAS, 317, 1014
\bibitem{}
Boyle B.J., Shanks T., Peterson B.A., 1988, MNRAS, 235, 935
\bibitem{}
Brandt W.N., et al., 2001, AJ, 122, 2810
\bibitem{}
Cattaneo A., 2001, MNRAS, 324, 128
\bibitem{}
Cattaneo A., 2002, MNRAS, 333, 353
\bibitem{}
Cattaneo A., Haehnelt M.G., Rees M.J., 1999, MNRAS, 308, 77
Cattaneo A., et al., 2003, in preparation
\bibitem{}
Chokshi A., Turner E.L., 1992, MNRAS, 259, 421
\bibitem{}
Cowie L.L., Barger A.J., Bautz M.W., Brandt W.N., Garmire G.P., 2003, ApJL, accepted, 
pre-print astro-ph/0301231
\bibitem{}
Daddi E., Broadhurst T., Zamorani, G., Cimatti A., R\"{o}ttgering H., Renzini A., 2001,
A\&A, 376, 825
\bibitem{}
Dietrich M., Appenzeller I., Hamann F., Heidt J., Jaeger K., Vestergaard M., Wagner S.J.,
2003, A\&A, 398, 891
\bibitem{}
Elvis M., et al., 1994, ApJS, 95, 1
\bibitem{}
Ferrarese L., Merritt D., 2000, ApJ, 529, 745
\bibitem{}
Franceschini A., Braito V., Fadda D., 2002, MNRAS, 335, 51p
\bibitem{}
Gebhardt K., et al., 2000a, AJ, 119, 1157
\bibitem{}
Gebhardt K., et al., 2000b, ApJ, 539, L13
\bibitem{}
Granato G.L., Silva L., Monaco P., Panuzzo P., Salucci P., De Zotti G.,
Danese L., 2001, MNRAS, 324, 757
\bibitem{}
Haehnelt M.G., Kauffmann G., 2000, MNRAS, 318, L35
\bibitem{}
Haehnelt M.G., Natarajan P., Rees M.J., 1998, MNRAS, 300, 817
\bibitem{}
Haiman Z., Hui L., 2001,  ApJ, 547, 27
\bibitem{}
Haiman Z., Loeb A., 1998, ApJ, 503, 505
\bibitem{}
Haiman Z., Menou K., 2000, ApJ, 531, 42
\bibitem{}
Hasinger G., 2002, in F. Jansen, ed., New Visions of the X-ray Universe in the
XMM-NEWTON and CHANDRA Era. ESTEC, the Netherlands, ESA SP-488
\bibitem{}
Hill G.J., Goodrich R.W., De Poy D.L., 1996, ApJ, 462, 163
\bibitem{}
Kauffmann G., Haehnelt M., 2000, MNRAS, 311, 576
\bibitem{}
Kauffmann G., Haehnelt M., 2002, MNRAS, 332, 529
\bibitem{}
La Franca F., Andreani P., Cristiani S., 1998, ApJ, 497, 529
\bibitem{}
Lawrence A., 1991, MNRAS, 252, 586
\bibitem{}
Lutz D., 2000, The Institut of Space and Astronautical Science Report, 14, 99
\bibitem{}
McLure R.J., Kukula M.J., Dunlop J.S., Baum S.A., O'Dea C.P., Hughes D.H.,
1999, MNRAS, 308, 377
\bibitem{}
Mainieri V., Bergeron J., Hasinger G., Lehmann I., Rosati P., Schmidt M.,
Szokoly G., Della Ceca R., 2002, A\&A, 393, 425
\bibitem{} 
Merritt D., Ferrarese L., 2001a, MNRAS, 320, L30
\bibitem{}
Merritt D., Ferrarese L., 2001b, ApJ, 547, 140
\bibitem{}
Nolan L.A., Dunlop J.S., Kukula M.J., Hughes D.H., Boroson T., Jimenez R.,
2001, MNRAS, 323, 308
\bibitem{}
Norman C., et al., 2002, ApJ, 571, 218
\bibitem{}
Romano D., Silva L., Matteucci F., Danese L., 2002, MNRAS, 334, 444
\bibitem{}
Rosati P., et al., 2002, ApJ, 566, 667
\bibitem{}
Rowan-Robinson M., 1995, MNRAS, 272, 737
\bibitem{}
Salucci P., Ratnam C., Monaco P., Danese L., 2000, MNRAS, 317, 488
\bibitem{}
Scheuer P.A.G., 1987, in T.J. Pearson, J.A. Zensus, eds., ``Superluminal Radio Sources''.
Cambridge University Press, Cambridge, p. 104
\bibitem{}
Sheth R. K., Bernardi M., Schechter P., et al., 2003, ApJ, submitted (astro-ph/0303092)
\bibitem{}
Shields G.A., et al., 2003, ApJ, accepted, pre-print astro-ph/0210050
\bibitem{}
Simpson C., 1998, MNRAS, 297, 39p
\bibitem{}
So{\l}tan A., 1982, MNRAS, 200, 115
\bibitem{}
Sturm E., Genzel R., Lutz D., Spoon H., Kunze D., Moorwood A., Netzer H., Sternberg A., 
1997, Bulletin of the American Astronomical Society, 29, 832
\bibitem{}
Thomas D., Maraston C., Bender R., 2002, Astrophysics and Space Science, 281, 371
\bibitem{}
Trager S.C., Faber S.M., Worthey G., Gonz\'alez J.J., 2000, AJ, 120, 165
\bibitem{}
Tremaine S., et al., 2002, 574, 540
\bibitem{}
Volonteri M., Haardt F., Madau P., 2002, Astrophysics and Space Science, 281, 501
\bibitem{}
Worthey G., 1994, ApJS, 95, 107
\bibitem{}
Yu Q., Tremaine S., 2002, MNRAS, 335, 965

\end{thebibliography}
\end{document}